\documentclass[final]{aipproc}
\layoutstyle{6x9}

\begin{document}                                              

\title{Cluster Model for Near-barrier Fusion Induced by Weakly Bound and Halo 
Nuclei}



\author{C.~Beck}{address={Institut Pluridisciplinaire Hubert Curien, UMR7178, 
CNRS-IN2P3 et Universit\'{e} Louis Pasteur (Strasbourg I), B.P. 28,
F-67037 Strasbourg Cedex 2, France}}

\author{N.~Keeley}{
   address={DSM/DAPNIA/SPhN CEA Saclay, F-91190 Gif-sur-Yvette, France}}

\author{A. Diaz-Torres}{
   address={The Australian National University, Canberra ACT 0200,
Australia}}

\begin{abstract}
The influence on the fusion process of coupling transfer/breakup channels 
is investigated for the medium weight $^{6,7}$Li+$^{59}$Co systems in the 
vicinity of the Coulomb barrier. Coupling effects are discussed within 
a comparison of predictions of the Continuum Discretized Coupled-Channels 
model. Applications to $^{6}$He+$^{59}$Co induced by the 
borromean halo nucleus $^{6}$He are also proposed.
\end{abstract}

\maketitle

\section{Introduction}

In reactions induced by light weakly bound nuclei, the influence on the fusion 
process of coupling to collective degrees of freedom and to breakup (BU) and 
transfer channels is a key point for a deeper understanding of few-body 
systems in quantum dynamics~\cite{Beck03,Diaz03,Beck06,Beck07a,Beck07b,Moro07}. 
Due to the very weak binding energies of halo nuclei, such as $^{6}$He,
a diffuse cloud of neutrons should lead to enhanced tunneling probabilities
below the Coulomb barrier as compared to predictions of one-dimensional barrier penetration
models~\cite{Beck03,Beck07a}. This was understood in terms of the dynamical 
processes arising from strong couplings to collective inelastic excitations of 
the target and projectile~\cite{Beck07a}. However, in the case of reactions 
where at least one of the colliding nuclei has a sufficiently low binding 
energy for BU to become a competitive process, conflicting model 
predictions and experimental results were reported~\cite{Beck07a}. Recent 
experimental results with $^{6,8}$He beams show that the halo of $^{6}$He does 
not enhance the fusion probability, confirming the prominent role of neutron 
transfers in $^{6}$He induced fusion reactions
\cite{Beck07a,Raabe04,Dipietro04,Navin04,DeYoung05,Penion07}. 

Excitation functions for sub- and near-barrier total - complete (CF) + 
incomplete (ICF) - fusion cross sections measured for the $^{6,7}$Li+$^{59}$Co 
reactions \cite{Beck03} when compared to Continuum-Discretized 
Coupled-Channels ({\sc CDCC}) calculations~\cite{Diaz03} indicate a small 
enhancement of total fusion (TF) for the more weakly bound $^{6}$Li below the 
Coulomb barrier, with similar cross sections for both reactions at and above 
the barrier \cite{Diaz03}. This result is consistent with BU - although with 
rather low cross sections even at incident energies larger than the 
Coulomb barrier~\cite{Beck06,Beck07a} - being more competitive for the 
$^{6}$Li+$^{59}$Co reaction.

In this contribution we present selected {\sc CDCC} calculations for elastic 
scattering, TF, transfer (TR), and BU of weakly bound stable ($^{6}$Li 
considered as a $\alpha$-$d$ cluster) and radioactive ($^{6}$He as a 
$\alpha$-2$n$ cluster) light projectiles from a medium-mass target ($^{59}$Co).  

\section{Full {\sc CDCC} description of $^{6}$Li+$^{59}$Co reaction}
      
      \begin{figure}
        \includegraphics[height=12cm]{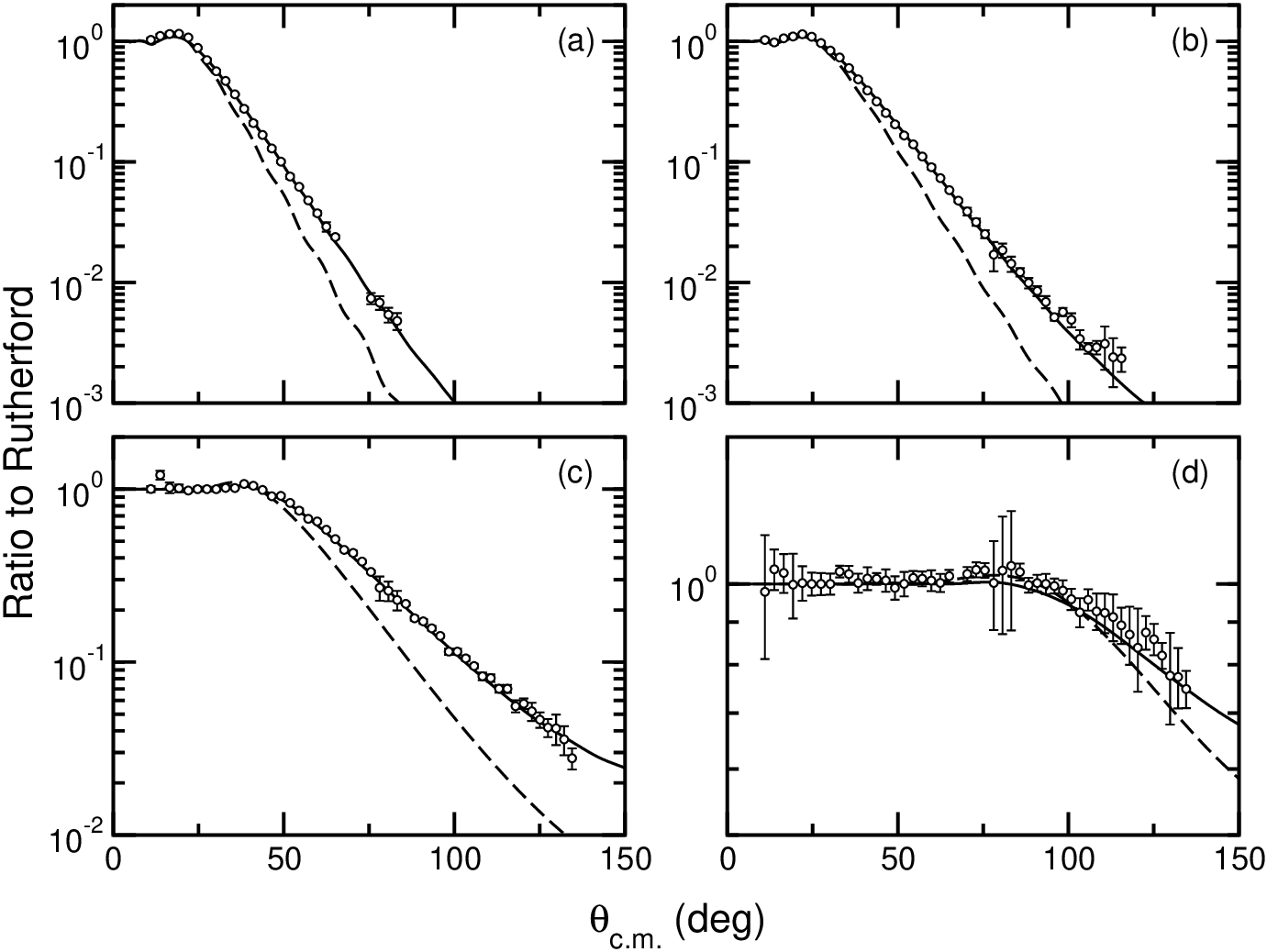}
        {\caption{\label{}{\small Elastic scattering for $^{6}$Li+$^{59}$Co
	                      at (a) 30 MeV, (b) 26 MeV, (c) 18 MeV and (d) 12 MeV
			      \cite{Beck06,Beck07a}. The curves correspond to 
			      {\sc CDCC} calculations \cite{Beck07b} with 
			      (solid lines) or without (dashed lines) couplings 
			      with the continuum as discussed in the text.}}}
       \end{figure}
       
In the present work, detailed {\sc CDCC} calculations for the interaction 
of $^{6}$Li on the medium-mass target $^{59}$Co are applied in order to 
provide a simultaneous description of elastic scattering, BU, TF as 
well as transfer. Details of the calculations concerning the breakup space 
(number of partial waves, resonances energies and widths, maximum continuum 
energy cutoff, potentials, ...) have been given in previous publications
\cite{Diaz03,Beck07b}. The {\sc CDCC} scheme is available in the general 
coupled channels code {\sc FRESCO} \cite{Diaz03}. Before investigating whether 
the proposed {\sc CDCC} formalism can be also applied to halo structures such 
as the borromean nucleu $^{6}$He, we present a complete description of the 
$^{6}$Li $\rightarrow$ $\alpha$+$d$ cluster as a two-body object. In earlier 
fusion calculations the imaginary parts of the off-diagonal couplings were 
neglected, while the diagonal couplings included imaginary parts~\cite{Diaz03}. 
We had used short-range imaginary fusion potentials for each fragment 
separately. This was equivalent to the use of incoming wave boundary conditions 
applied in previous {\sc CCFULL} calculations~\cite{Beck03}. Here full 
continuum couplings have been taken into account so as to reproduce the 
elastic scattering data~\cite{Beck06,Beck07a}.

Results of the comparison of the {\sc CDCC} calculations for the elastic
scattering with data of Ref.~\cite{Beck06,Beck07a} are shown in Fig.~1 
for $^{6}$Li+$^{59}$Co at four different incident energies. The two different 
curves are the results of calculations performed with (solid lines) and 
without (dashed lines) $^{6}$Li $\rightarrow$ $\alpha$ + $d$ breakup couplings 
The agreement between the full calculations and data is very good. The effect 
of the BU channel on elastic scattering is illustrated by the difference between 
one-channel calculations and full {\sc CDCC} results. 

       \begin{figure}
       \includegraphics[height=12cm]{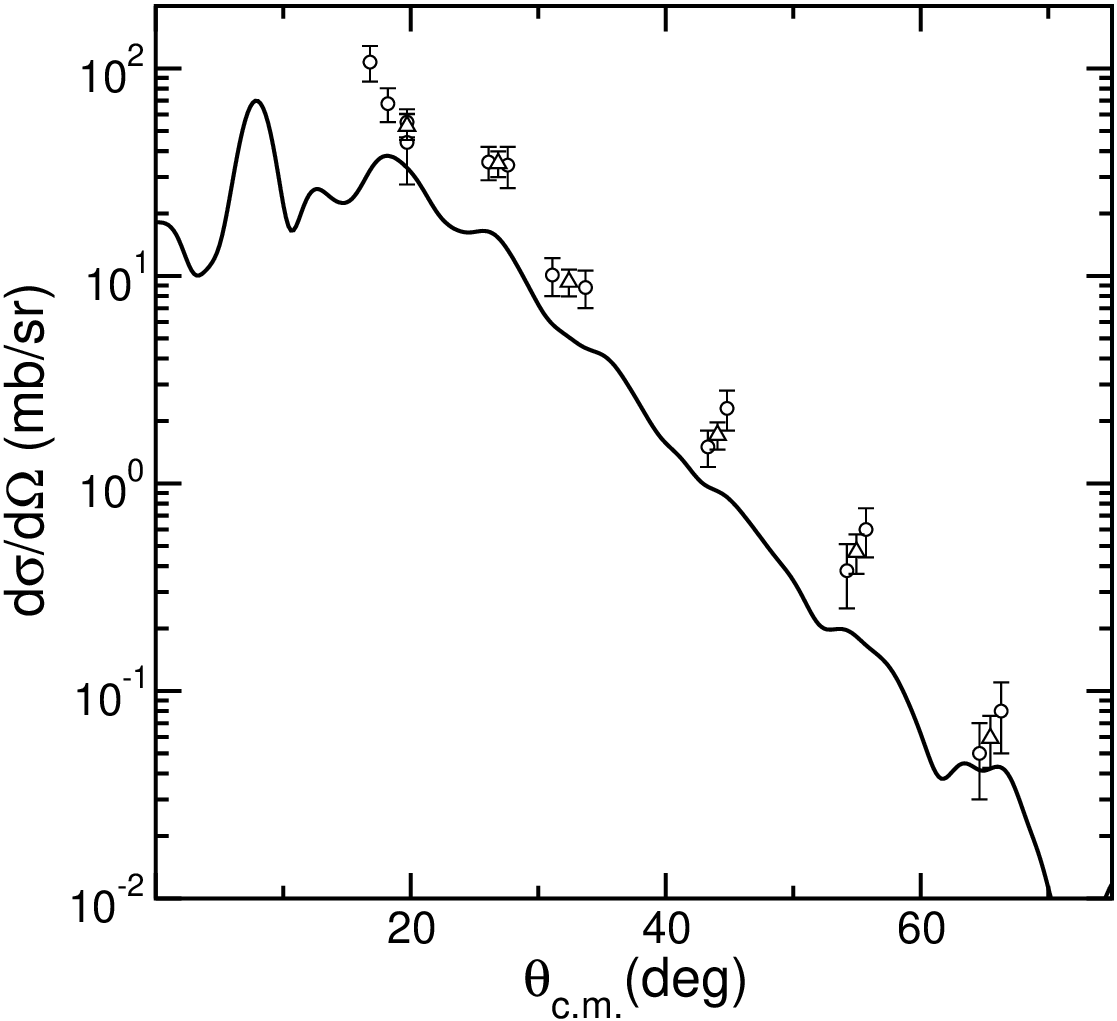}
      {\caption{\label{}{\small CDCC calculation for the angular distribution 
                                of the $^6$Li $\rightarrow$ $\alpha + d$ 
				sequential breakup via the 2.18 MeV $3^+$ state 
                                of $^6$Li compared to data of 
				$^{6}$Li+$^{59}$Co reaction at 41 MeV.}}}
        \end{figure}

The total calculated BU cross sections were obtained by integrating 
contributions from the states in the continuum up to 8 MeV. They are found 
to be negligible fractions (between 3.7--9.7 \%) of the total reaction 
cross sections and small compared with the TF cross sections (both the
data~\cite{Beck03} and the two different sets of {\sc CDCC} calculations
\cite{Diaz03,Beck07b}). This conclusion can be verified by the angular 
correlations for the sequential BU of $^6$Li via the 2.18 MeV $3^+$ excited 
state \cite{Beck07b} plotted in Fig.~2 at 41 MeV. Sequential BU via this state 
is the dominant contribution to the total $^6$Li $\rightarrow$ $\alpha + d$ 
breakup cross section ($\sigma_{tot}$ $\approx$ 80 mb). Its {\sc CDCC} cross 
section ($\sigma_{seq}$ = 22.5 mb) is, however, significantly smaller than the 
report experimental value ($\sigma_{exp}$ = $45 \pm 10$ mb)~\cite{Beck07b}.
       
       \begin{figure}
       \includegraphics[height=10cm]{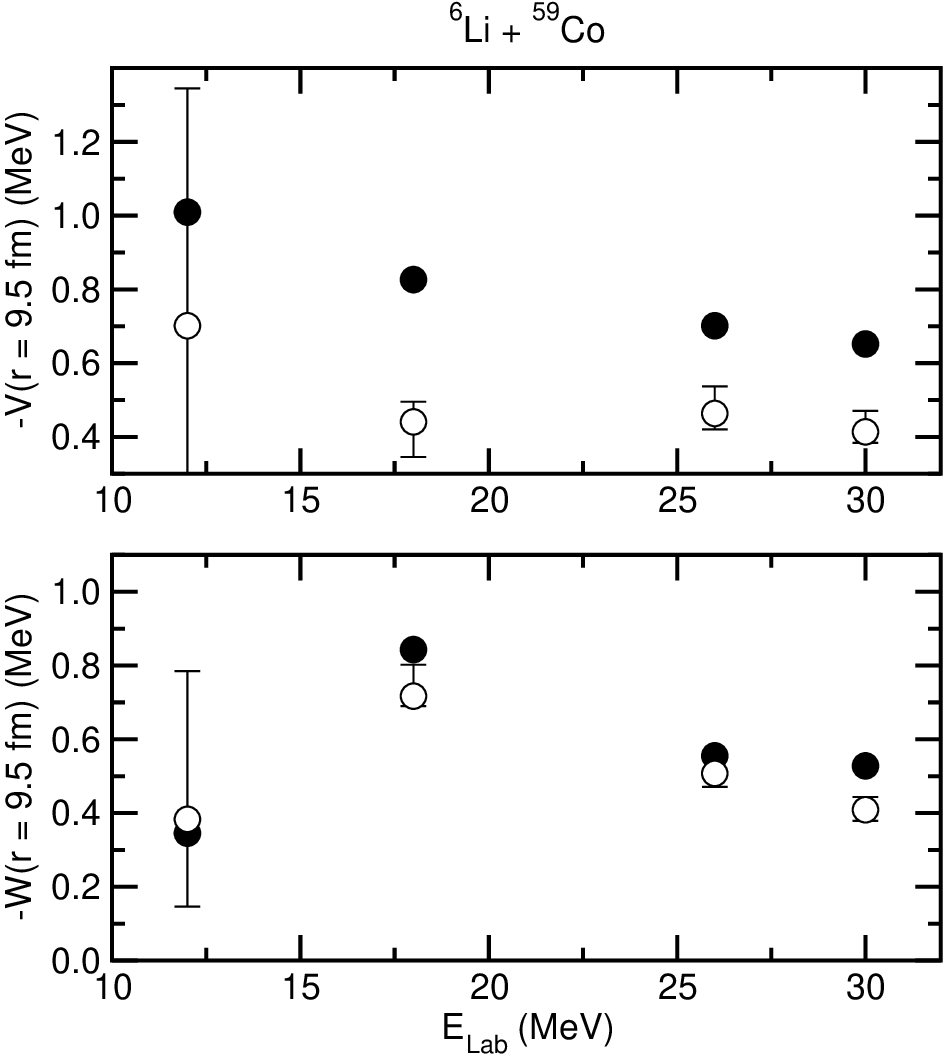}
                      {\caption{\label{}{\small Energy dependence of the real 
		      and imaginary parts of the ``bare + DPP'' potentials 
		      as generated by the CDCC calculations (filled circles)
                      and the best OM fits potentials (open circles)
                      for the $^6$Li+$^{59}$Co system at a radial distance of 
		      r = 9.5 fm.}}}
        \end{figure}
	
The total reaction cross sections obtained for $^{6,7}$Li+$^{59}$Co either 
from fits with Optical Model (OM) potentials~\cite{Beck07b} or {\sc CDCC} 
calculations confirm the observed small enhancement of TF cross section for 
the more weakly bound $^{6}$Li nucleus at sub-barrier energies~\cite{Beck03}. 

Fig.~3 illustrates how the surface strengths of the ``bare plus dynamic 
polarization (DPP)'' potentials, produced by the coupling of the BU channel, 
are in good agreement with OM potentials. Apparently, they exhibit the energy 
dependence characteristic of the ``threshold anomaly'' (TA), i.e.\ a rise in 
the strength of the real part as the incident energy is reduced towards the 
Coulomb barrier accompanied by a drop in that of the imaginary part. However, 
this conclusion largely rests on the values at 12 MeV incident energy, and as 
can be seen from the error bars, the potential strength in the nuclear surface 
is effectively not determined by the data due to its rather poor precision, a 
very wide range of values giving equally good fits to the data. The spread in 
values for the other energies, while much less than that at 12 MeV, is still 
such that we are unable to draw any concrete conclusions concerning the 
presence or absence of a TA in $^6$Li+$^{59}$Co. Similar observations have 
been made for $^7$Li+$^{59}$Co and the question of the occurence a ``BU 
threshold anomaly'' (BTA)~\cite{Beck07b} for lithium projectiles on light- 
and medium-mass targets remains widely open. More precise experimental elastic 
scattering data at sub- and near-barrier energies will, therefore, be 
necessary.

\section{Predictions for $^{6}$He+$^{59}$Co fusion and conclusions}

Calculations applied to the two-neutron halo nucleus $^{6}$He is much more 
complicated since $^{6}$He breaks into three fragments ($\alpha$+n+n) 
instead of two ($\alpha$+d), and the {\sc CDCC} method for two-nucleon halo 
nuclei has not yet been implemented in FRESCO~\cite{Diaz03}. A dineutron model 
is adopted for the $^{6}$He+$^{59}$Co reaction~\cite{Beck06,Beck07a,Beck07b}: 
i.e. we assume a two-body cluster structure of $^{6}$He = $^{4}$He+$^{2}$n with 
an $\alpha$ particle core coupled to a single particle representing a di-neutron 
($^{2}$n) like cluster. Couplings to resonant (2$^{+}$, E$_{ex}$ = 0.826 MeV) 
and non-resonant continuum states (up to f-waves) are included. The fact that 
the dineutron is not an object with both fixed size and fixed energy 
(Heisenberg principle) might be a critical point in the present model. Results 
of the {\sc CDCC} calculations for TF of the $^{6}$He+$^{59}$Co system were 
compared to $^{4}$He+$^{59}$Co and $^{6}$Li+$^{59}$Co, used as basis reactions 
(see Refs.\cite{Beck06,Beck07a,Beck07b}). We observed that calculations with 
and without breakup give much larger TF cross sections for $^{6}$He compared to 
$^{4}$He and $^{6}$Li. The inclusion of the couplings to the BU channels 
notably increases the TF cross section for all energies. The predictions for 
the $^{59}$Co target somewhat over predict the data published for other 
medium-mass targets such as $^{64}$Zn~\cite{Dipietro04} and 
$^{63,65}$Cu~\cite{Navin04}. Extended calculations are in progress to quantify 
the role of 1n- and 2n-transfer channels found to be significant in recent 
$^{6}$He data~\cite{Raabe04,Dipietro04,Navin04,DeYoung05,Penion07}. 

The {\sc CDCC} method~\cite{Diaz03} can be used to provide the almost complete 
theoretical description of all competing processes (elastic scattering, TF, TR 
and BU) in a consistent way. In this contribution we have shown that the 
$^{6}$Li+$^{59}$Co reaction can be fairly well understood in this framework 
although {\sc CDCC} does not separate CF from ICF. {\sc CDCC} results for 
the $^{6}$He+$^{59}$Co fusion process are also briefly discussed. A complete 
understanding of the reaction dynamics involving couplings to the BU and the 
neutron transfer channels will need high-intensity radioactive ion beams to 
permit measurements at deep sub-barrier energies and precise measurements of 
elastic scattering (search for TA and BTA) and yields leading to TR channels 
and to BU itself. The application of four-body (required for an accurate 
$\alpha$-n-n description of $^{6}$He) {\sc CDCC} models still under current 
development \cite{Moro07} will then be highly desirable. It would be 
interesting to see what difference this more accurate model would have on the 
BU coupling effect on TF if applied to a fusion calculation in a similar 
manner to the calculations presented in this contribution.

\bibliographystyle{aippprocl}


%


\end{document}